# Influence of electron-hole drag on conductivity of neutral and gated graphene

I. I. Boiko[1]

*Institute of Semiconductor Physics, NAS of Ukraine, 45, pr. Nauky, 03028, Kiev, Ukraine*

(Dated: November 4, 2010)

Conductivity of monolayer and two-layer graphene is considered with due regard for mutual drag of band electrons and holes. Search of contribution of the drag in conductivity shows that this effect can sufficiently influence on mobility of carriers, which belong to different groups and have different drift velocities. In two-layer system the mutual drag can even change the direction of partial current in separate layer.

## Introduction

The great interest now exists to systems of 2D-carriers with Dirac-like dispersion law (see Refs. [1] – [4]):

$$\varepsilon^{(e,h)}(\vec{k}_\perp) = \hbar v_F k_\perp . \qquad (I.1)$$

Physical properties of such systems are substantially distinct on that for the systems with parabolic dispersion law. The most surprising point is real possibility to realize a substantially two-dimensional system at elevated temperature.

In general case conductivity of graphene is bipolar. For neutral graphene Fermi-level $\varepsilon_F$ lays exactly in the point $\varepsilon = 0$; so hole and electron densities are equal:

$$n_e = n_h = \pi^{-2}\int d^2\vec{k}_\perp [1+\exp(\varepsilon(k_\perp)/k_B T)]^{-1} = (2/\pi)(k_B T/\hbar v_F)^2 \int_0^\infty t[1+\exp(t)]^{-1}dt \approx 0.524(k_B T/\hbar v_F)^2 . \quad (I.2)$$

It follows from here at $v_F = 10^8 cm/s$: $n_{e,h}(T=1K)) \approx 7.2 \cdot 10^5 cm^{-2}$. Therefore, neutral graphene can be considered as a semimetal with extremely small number of mobile carriers.

Microscopic velocities of electrons and holes are equal also:

$$\vec{v}^{(e,h)}(\vec{k}_\perp) = \frac{1}{\hbar}\frac{\partial \varepsilon^{(e,h)}(\vec{k}_\perp)}{\partial \vec{k}_\perp} = v_F \frac{\partial k_\perp}{\partial \vec{k}_\perp} = v_F \frac{\vec{k}_\perp}{k_\perp} . \qquad (I.3)$$

If Fermi-level is shifted by applying of some bias voltage $V$, to a gate, we obtain *n*-graphene or *p*-graphene with the density of basic carriers ($a = e$ or $h$)

$$n_a = \frac{1}{\pi^2}\int d^2\vec{k}_\perp \left[1+\exp\left(\frac{\varepsilon^{(a)}(k_\perp)-eV_a}{k_B T}\right)\right]^{-1} = \left(\frac{k_B T}{v_F \hbar}\right)^2 N(\kappa_a) . \qquad (I.4)$$

Here $\kappa_a = eV_a/k_B T > 0$; $N(\kappa_a) = (2/\pi)\int_0^\infty \kappa[1+\exp(\kappa-\kappa_a)]^{-1}d\kappa$; $N(0) = 0.524$. At $\kappa_a >> 1$ (high level of degeneracy) $n_a = \pi^{-1}(eV_a/v_F \hbar)^2$. Let $V_a = 10V$; then $n_a = 1.5 \cdot 10^{16} cm^{-2}$. This density is rather high in comparison with the density of carriers in neutral graphene.

---

[1] Electron address:  igorboiko@yandex.ru

## 1. Kinetic equation

### *1. 1. General form of quantum kinetic equation*

Kinetic equation for the density matrix of carriers of *a*-type at arbitrary quantum numbers presentation has the form (see. Refs. [5] and [6])

$$i\hbar \frac{\partial f_{AB}^{(a)}(t)}{\partial t} = e \sum_{\Gamma} \{\varphi_{A\Gamma}^{(E)} f_{\Gamma B}^{(a)}(t) - f_{A\Gamma}^{(a)}(t) \varphi_{\Gamma B}^{(E)}\} + i\hbar \operatorname{St} f_{AB}^{(a)}(t) \ . \quad (1.1)$$

In uniform space the density matrix is diagonal one: $f_{AB}^{(a)} = \delta_{AB} f_{AA}^{(a)} \equiv \delta_{AB} f_A^{(a)}$. For applied electrical field $\vec{E}$ the potential $\varphi^{(E)}(\vec{r}) = -\vec{E}\,\vec{r}$.

The total collision integral $\operatorname{St} f_A^{(a)}$ for mobile *a*-carriers in *A*-state is the sum of collision integrals concerning as external scattering system (*S*) as mobile carriers of other types:

$$\operatorname{St} f_A^{(a)} = \operatorname{St}_{a-S} f_A^{(a)} f_A + \sum_b \operatorname{St}_{a-b} f_A^{(a)} \ . \quad (1.2)$$

Here

$$\operatorname{St}_{a-S} f_A^{(a)} = -\frac{e^2}{(2\pi)^3 \hbar^2} \sum_B \int d^3\vec{q} \int d\omega \, \delta(\omega - \omega_{AB}^{(a)}) |(b_{\vec{q}})_{AB}^{(a)}|^2 \langle \varphi_{(S)}^2 \rangle_{\omega, \vec{q}} \times$$
$$\times \left\{ f_A^{(a)} - f_B^{(a)} + [f_A^{(a)}(1 - f_B^{(a)}) + f_B^{(a)}(1 - f_A^{(a)})]\tanh\left(\frac{\hbar\omega}{2k_B T}\right)\right\} \ ; \quad (1.3)$$

$$\operatorname{St}_{a-b} f_A^{(a)} = \frac{2\pi}{\hbar^2} \sum_{BA'B'} [f_B^{(a)}(1 - f_A^{(a)}) f_{A'}^{(b)}(1 - f_{B'}^{(b)}) - f_{B'}^{(b)}(1 - f_{A'}^{(b)}) f_A^{(a)}(1 - f_B^{(a)})] \delta(\omega_{AB}^{(a)} - \omega_{A'B'}^{(b)}) |\tilde{V}_{ABB'A'}^{(a,b)}(\omega_{AB})|^2. \quad (1.4)$$

The value $\varphi_{(S)}$ is a screened potential for external scattering system (impurities, phonons *et al*);

$$\tilde{V}_{ABB'A'}^{(a,b)}(\omega) = (2\pi)^{-3} \int (b_{\vec{q}})_{AB}^{(a)} (b_{-\vec{q}})_{B'A'}^{(b)} [4\pi e^2 / q^2 \varepsilon(\omega, \vec{q})] d^3\vec{q} \ ;$$

$$(b_{\vec{q}})_{AB\,AB}^{(a)} = \int [\Psi_A^{(a)}(\vec{r})]^* \exp(i\vec{q}\vec{r}) \Psi_B^{(a)}(\vec{r}) d^3\vec{r} \ ; \quad \omega_{AB}^{(a)} = (1/\hbar)(\varepsilon_A^{(a)} - \varepsilon_B^{(a)}) ; \quad (1.5)$$

$\varepsilon_A^{(a)}$ is the energy of the band *a*-particle in *A*-state. Spin is not discussed in this paper.

### *1.2. Two-dimensional quantum kinetic equation in $\vec{k}_\perp$-representation*

In uniform two-dimensional space the natural basis is set of plane waves, and corresponding quantum numbers are wave-vectors $\vec{k}_\perp = \vec{e}_x k_x + \vec{e}_y k_y$. Assume (for band particles of arbitrary type)

$$\Psi_A(\vec{r}) \to L^{-1} \exp(i\vec{k}_\perp \vec{r}_\perp) \delta_{z,0} \ .$$

Then (see Eq. (1.5)) $|(b_{\vec{q}})_{AB}|^2 \to (2\pi/L)^2 \delta(\vec{k}_\perp - \vec{k}_\perp' - \vec{q}_\perp)$. Here *L* is size of crystal for any orthogonal direction in the plane $z = 0$.

For applied uniform electrical field $\vec{E}_\perp$, laying in *x y* - plane, $\varphi^{(E)}(\vec{r}_\perp) = -\vec{E}_\perp \vec{r}_\perp$. The matrix element of this potential is

$$\varphi_{\vec{k}\perp,\vec{k}'\perp}^{(E)} = -\vec{E}_\perp L^{-2} \iint_{LL} \vec{r}_\perp \exp[i(\vec{k}_\perp' - \vec{k}_\perp)] d^2\vec{r}_\perp = -i\vec{E}_\perp \partial \delta_{\vec{k}_\perp,\vec{k}'\perp} / \partial \vec{k}_\perp. \ .$$

It follows from here:

$$\sum_{\vec{k}'}\{[\varphi^{(E)}(\vec{r})]_{\vec{k},\vec{k}'}f_{\vec{k}',\vec{k}} - f_{\vec{k},\vec{k}}[\varphi^{(E)}(\vec{r})]_{\vec{k}',\vec{k}}\} = -i\vec{E}\frac{\partial f_{\vec{k}_\perp}}{\partial \vec{k}_\perp}.$$

As total, the quantum kinetic equation (1.1) takes the form of integro-differential equation. For stationary case

$$\frac{e_a}{\hbar}\vec{E}_\perp \frac{\partial f^{(a)}_{\vec{k}_\perp}}{\partial \vec{k}_\perp} = \text{St } f^{(a)}_{\vec{k}_\perp}. \tag{1.6}$$

Here and in the following equations the symbols $a$, $b = e$ or $h$; $e_e = -e < 0$ and $e_h = e > 0$.

Write the collision integral in Eq. (1.2) as the sum

$$\text{St} f^{(a)}_{\vec{k}_\perp} = \text{St}_{a-S} f^{(a)}_{\vec{k}_\perp} + \sum_b \text{St}_{a-b} f^{(b)}_{\vec{k}_\perp}, \tag{1.7}$$

where (see Eqs. (1.3) and (1.4))

$$\text{St}_{a-S} f^{(a)}_{\vec{k}_\perp} = -\frac{e^2}{(2\pi)^3 \hbar}\int d\omega \delta(\hbar\omega - \varepsilon^{(a)}_{\vec{k}_\perp} + \varepsilon^{(a)}_{\vec{k}_\perp-\vec{q}_\perp}) \times$$
$$\times \int\left\{\left[f^{(a)}_{\vec{k}_\perp}(1-f^{(a)}_{\vec{k}_\perp-\vec{q}_\perp}) + f^{(a)}_{\vec{k}_\perp-\vec{q}_\perp}(1-f^{(a)}_{\vec{k}_\perp})\right]\tanh\left(\frac{\hbar\omega}{2k_B T}\right) + f^{(a)}_{\vec{k}_\perp} - f^{(a)}_{\vec{k}_\perp-\vec{q}_\perp}\right\}\langle\varphi^2_{(S)}\rangle_{\omega,\vec{q}} d^3\vec{q}; \tag{1.8}$$

$$\text{St}_{a-b} f^{(a)}_{\vec{k}_\perp} = \frac{e^4}{2\pi \hbar\varepsilon_L^2}\int d^2\vec{k}'_\perp \int \frac{d^2\vec{q}_\perp}{q_\perp^2}(\varepsilon^{(a)}_{\vec{k}_\perp} - \varepsilon^{(a)}_{\vec{k}_\perp-\vec{q}_\perp} + \varepsilon^{(b)}_{\vec{k}'_\perp-\vec{q}_\perp} - \varepsilon^{(b)}_{\vec{k}'_\perp}) \times$$
$$\times [f^{(a)}_{\vec{k}_\perp-\vec{q}_\perp}(1-f^{(a)}_{\vec{k}_\perp})f^{(b)}_{\vec{k}'_\perp}(1-f^{(b)}_{\vec{k}'_\perp-\vec{q}_\perp}) - f^{(b)}_{\vec{k}'_\perp-\vec{q}_\perp}(1-f^{(b)}_{\vec{k}'_\perp})f^{(a)}_{\vec{k}_\perp}(1-f^{(a)}_{\vec{k}_\perp-\vec{q}_\perp})]. \tag{1.9}$$

In these expressions the contribution of two-dimensional band carriers in screening is considered as negligible in comparison with dielectric constant of crystal lattice $\varepsilon_L$.

## 2. Balance equations

Apply to both sides of Eq. (1.6) the operator

$$\pi^{-2}\int \vec{k}_\perp d^2\vec{k}_\perp. \tag{2.1}$$

The expression, obtained by this way, is exact first momentum of quantum kinetic equation in space of plane wave vectors. It has the meaning of a balance equation for dynamic and dissipative forces, applied to the band particles from *a*-group:

$$e_a\vec{E}_\perp + \vec{F}^{(a,S)}_\perp + \sum_b \vec{F}^{(a,b)}_\perp = 0. \tag{2.2}$$

Here

$$\vec{F}^{(a,S)}_\perp = \frac{\hbar}{\pi^2 n_a}\int \vec{k}_\perp [\text{St}^{(a,S)} f^{(a)}_{\vec{k}_\perp}] d^2\vec{k}_\perp =$$
$$= -\frac{2e^2}{(2\pi)^4 n_a}\int d^2\vec{k}_\perp \int \vec{q}_\perp d^2\vec{q}_\perp \int d\omega \delta(\hbar\omega - \varepsilon^{(a)}_{\vec{k}_\perp} + \varepsilon^{(a)}_{\vec{k}_\perp-\vec{q}_\perp})\langle\varphi^2_{(S)}\rangle_{\omega,\vec{q}_\perp} \times$$
$$\times \left\{\left[f^{(a)}_{\vec{k}_\perp}(1-f^{(a)}_{\vec{k}_\perp-\vec{q}_\perp}) + f^{(a)}_{\vec{k}_\perp-\vec{q}_\perp}(1-f^{(a)}_{\vec{k}_\perp})\right]\tanh\left(\frac{\hbar\omega}{2k_B T}\right) + f^{(a)}_{\vec{k}_\perp} - f^{(a)}_{\vec{k}_\perp-\vec{q}_\perp}\right\}; \tag{2.3}$$

$$\vec{F}^{(a,b)}_\perp = \frac{\hbar}{\pi^2 n_a}\int \vec{k}_\perp [\text{St}^{(a,b)} f^{(a)}_{\vec{k}_\perp}] d^2\vec{k}_\perp =$$

$$= \frac{2e^4\hbar}{(2\pi)^3 \varepsilon_L^2 n_a} \int d^2\vec{k}_\perp \int d^2\vec{k}'_\perp \int d\omega \int \vec{q}_\perp \frac{d^2\vec{q}_\perp}{q_\perp^2} \delta(\hbar\omega - \varepsilon^{(a)}_{\vec{k}_\perp} + \varepsilon^{(a)}_{\vec{k}_\perp - \vec{q}_\perp}) \delta(\hbar\omega - \varepsilon^{(b)}_{\vec{k}'_\perp} + \varepsilon^{(b)}_{\vec{k}'_\perp - \vec{q}_\perp}) \times \quad (2.4)$$

$$\times [f^{(a)}_{\vec{k}_\perp - \vec{q}_\perp}(1 - f^{(a)}_{\vec{k}_\perp}) f^{(b)}_{\vec{k}'_\perp}(1 - f^{(b)}_{\vec{k}'_\perp - \vec{q}_\perp}) - f^{(b)}_{\vec{k}'_\perp - \vec{q}_\perp}(1 - f^{(b)}_{\vec{k}'_\perp}) f^{(a)}_{\vec{k}_\perp}(1 - f^{(a)}_{\vec{k}_\perp - \vec{q}_\perp})].$$

In Eq. (2.2) the first term is the dynamical force, acting on charged carriers from the side of applied electrical field, the second term is the friction force from the side of an external scattering system, the last term is the summarized friction force, concerning other band carriers and applied to the mobile carriers of $a$-group. As far as $\vec{F}_\perp^{(a,a)} = 0$, one can consider $\vec{F}_\perp^{(a,b)}$ as a force of the inter-group drag.

Farther we use the model nonequilibrium distribution functions (see Refs. [5], [6]):

$$f^{(a)}_{\vec{k}_\perp} = \frac{1}{1 + \exp\{[\hbar \vec{k}_\perp (\vec{v}^{(a)}_{\vec{k}_\perp} - \vec{u}^{(a)}_\perp) - eV_a]/k_B T\}} = \frac{1}{1 + \exp\{[\hbar(k_\perp v_F - \vec{k}_\perp \vec{u}^{(a)}_\perp) - eV_a]/k_B T\}} . \quad (2.5)$$

Here $V_a$ is transverse bias voltage applied to the graphene layer with the help of some gate system (see, for instance, Fig. 1). The function $f^{(a)}_{\vec{k}_\perp}$ contains a vector $\vec{u}^{(a)}_\perp$ as the model parameter. The meaning of that becomes evident after calculation of the averaged velocity of $a$-particles:

$$\langle \vec{v}^{(a)}_{\vec{k}_\perp} \rangle = (\pi^2 n_a)^{-1} \int \vec{v}^{(a)}_{\vec{k}_\perp} f^{(a)}_{\vec{k}_\perp} d^2\vec{k}_\perp = \vec{u}^{(a)}_\perp . \quad (2.6)$$

Thus, the vector $\vec{u}^{(a)}_\perp$ is the drift velocity for the group of $a$-particles. The total density of a current in bipolar graphene is

$$\vec{j} = - e n_e \vec{u}^{(e)}_\perp + e n_h \vec{u}^{(h)}_\perp . \quad (2.7)$$

Substituting the model functions (2.5) in the formulae (2.3), (2.4) and carrying out the linearization of the forces $\vec{F}_\perp^{(a,S)}$ and $\vec{F}_\perp^{(a,b)}$ over the drift velocities $\vec{u}^{(a)}_\perp$ and $\vec{u}^{(b)}_\perp$, we come to the following convenient forms:

$$\vec{F}_\perp^{(a,S)} = - e_a \beta^{(a)} \vec{u}^{(a)}_\perp \quad ; \quad \vec{F}_\perp^{(a,b)} = - e_a \xi^{(a,b)} (\vec{u}^{(a)}_\perp - \vec{u}^{(b)}_\perp) . \quad (2.8)$$

$$\xi^{(a,b)} = \frac{e^2 e_a \hbar^2}{(2\pi)^3 k_B T \varepsilon_L^2 n_a} \int d^2\vec{k}_\perp \int d^2\vec{k}'_\perp \int d^2\vec{q}_\perp \int d\omega \, \delta(\hbar\omega - \varepsilon^{(a)}_{\vec{k}_\perp} + \varepsilon^{(a)}_{\vec{k}_\perp - \vec{q}_\perp}) \delta(\hbar\omega - \varepsilon^{(b)}_{\vec{k}'_\perp} + \varepsilon^{(b)}_{\vec{k}'_\perp - \vec{q}_\perp}) \times \quad (2.9)$$

$$\times (f_0(\varepsilon^{(a)}_{\vec{k}_\perp}) - f_0(\varepsilon^{(a)}_{\vec{k}_\perp - \vec{q}_\perp}))(f_0(\varepsilon^{(b)}_{\vec{k}'_\perp}) - f_0(\varepsilon^{(b)}_{\vec{k}'_\perp - \vec{q}_\perp})) \sinh^{-2}(\hbar\omega/2k_B T) ;$$

$$\beta^{(a)} = \frac{2 e_a \hbar}{(2\pi)^4 k_B T n_a} \int d^2\vec{k}_\perp \int q_\perp^2 d^2\vec{q} \int d\omega \, \delta(\hbar\omega - \varepsilon^{(a)}_{\vec{k}_\perp} + \varepsilon^{(a)}_{\vec{k}_\perp - \vec{q}_\perp}) \times \quad (2.10)$$

$$\times (f_0(\varepsilon^{(a)}_{\vec{k}_\perp}) - f_0(\varepsilon^{(a)}_{\vec{k}_\perp - \vec{q}_\perp})) \sinh^{-1}(\hbar\omega/k_B T) \langle \varphi^2_{(S)} \rangle_{\omega, \vec{q}_\perp} .$$

Note, that

$$e_a n_a \xi^{(a,b)} = e_b n_b \xi^{(b,a)} . \quad (2.11)$$

For elastic and isotropic scattering

$$\langle \varphi^2 \rangle_{\omega, \vec{q}_\perp} = \langle \varphi^2 \rangle_{q_\perp} \delta(\omega) ; \quad (2.12)$$

$$\beta^{(a)} = - \frac{e_a \hbar}{(2\pi)^4 n_a} \int d^2\vec{k}_\perp \int q_\perp^2 d^2\vec{q}_\perp \, \delta(\varepsilon^{(a)}_{\vec{k}_\perp} - \varepsilon^{(a)}_{\vec{k}_\perp - \vec{q}_\perp}) \frac{\partial f_0(\varepsilon^{(a)}_{\vec{k}_\perp})}{\partial \varepsilon^{(a)}_{\vec{k}_\perp}} \langle \varphi^2_{(S)} \rangle_{q_\perp} . \quad (2.13)$$

If the considered carriers and two-dimensional scattering system are separated in space on some distance $l$, one has to introduce in Eqs. (2.3), (2.4), (2.9), (2.10) and (2.13) the factor $\exp(-2q_\perp l)$ under the sign of integral over $\vec{q}_\perp$.

Combining Eqs. (2.2) and (2.8), we obtain the system of vector equations (here $\xi = -\xi^{(e,h)} > 0$):

$$\vec{E}_\perp - \beta^{(e)} \vec{u}^{(e)} + \xi(\vec{u}^{(e)} - \vec{u}^{(h)}) = 0 ; \qquad (2.14)$$

$$\vec{E}_\perp - \beta^{(h)} \vec{u}^{(h)} - \xi(n_h/n_e)(\vec{u}^{(h)} - \vec{u}^{(e)}) = 0 . \qquad (2.15)$$

One can see from here that in absence of a mutual drag ($\xi = 0$) the values $(-\beta^{(e)})$ and $\beta^{(h)}$ are reverse mobilities of electrons and holes. In the last two equations the terms, containing the factor $\xi$, represent the electron-hole drag directly.

### 3. Interaction with scattering system

To calculate the values $\beta^{(a)}$ we have at first to construct the correlator $\langle \varphi^2_{(S)} \rangle_{\vec{q}_\perp}$.

#### *3.1. Neutral impurities*

For external scattering system represented by neutral impurities, arranged in a graphene plane with surface concentration $n_{NI}^{(2)}$ (see below Eq. (*A*.3) and above Eqs. (2.5), (2.13)),

$$\langle \varphi^2_{NI} \rangle_{q_\perp} = 9\pi e^2 n_{NI}^{(2)} / 2\varepsilon_L^2 q_B^2 ; \qquad (3.1)$$

$$\left|\beta^{(a)}_{(NI)}\right| = \frac{9e^3 \hbar n_{NI}^{(2)}}{4\pi\varepsilon_L^2 q_B^2 n_a (k_B T)^2} \left(\frac{k_B T}{v_F \hbar}\right)^6 \int_0^\infty \frac{\kappa^4 \exp(\kappa - \kappa_a)}{[1 + \exp(\kappa - \kappa_a)]^2} d\kappa = \frac{e^3 n_{NI}^{(2)} (k_B T)^2}{\varepsilon_L^2 q_B^2 v_F^4 \hbar^3} \Phi(\kappa_a) ; \qquad (3.2)$$

$$\Phi(\kappa_a) = \frac{9}{8} \frac{\int_0^\infty \kappa^4 \exp(\kappa - \kappa_a) [1 + \exp(\kappa - \kappa_a)]^{-2} d\kappa}{\int_0^\infty \kappa [1 + \exp(\kappa - \kappa_a)]^{-1} d\kappa} ; \quad \Phi(0) \approx 31.1; \quad \Phi(\kappa_a \gg 1) \approx \frac{9\kappa_a^2}{4} = \left(\frac{3eV_a}{2k_B T}\right)^2 . \qquad (3.3)$$

Therefore, for neutral graphene ($V_a = 0$)

$$\left|\beta^{(a)}_{(NI)}\right| = 31 \frac{e^3 n_{NI}^{(2)} (k_B T)^2}{\varepsilon_L^2 q_B^2 v_F^4 \hbar^3} . \qquad (a = e, h) \qquad (3.4)$$

Thus the value $\beta^{(a)}_{(NI)}$ is proportional to the square of temperature.

For monopolar graphene, where $eV_a / k_B T \gg 1$, the mobility is inverse proportional to the square of gate voltage:

$$\mu^{(a)}_{(NI)} = \left|\beta^{(a)}_{(NI)}\right|^{-1} = \frac{4\varepsilon_L^2 q_B^2 v_F^4 \hbar^3}{9e^3 n_{NI}^{(2)} (eV_a)^2} . \qquad (a = e, h) \qquad (3.5)$$

If neutral impurities are arranged in a plane, separated of graphene plane on the distance $l$, where $l \cdot \max\{k_B T, eV_a\} > \hbar v_F$, then

$$\left|\beta^{(a)}_{(NI)}\right| = \frac{9e^3 n_{NI}^{(2)}}{64\pi\varepsilon_L^2 q_B^2 k_B T v_F l^3} \Lambda(\kappa_a) ; \quad \Lambda(\kappa_a) = \frac{\int_0^\infty \kappa \exp(\kappa - \kappa_a)[1 + \exp(\kappa - \kappa_a)]^{-2} d\kappa}{\int_0^\infty \kappa [1 + \exp(\kappa - \kappa_a)]^{-1} d\kappa} . \qquad (3.6)$$

Here $\Lambda(0) = 0.843$ ; $\Lambda(\kappa_a \gg 1) = 2/\kappa_a$.

### 3.2. Charged impurities

For charged impurities, arranged in graphene plane (see Eqs. (*A*4) and (2.13)),

$$\langle \varphi^2_{CI} \rangle_{\vec{q}\perp} = \frac{(2\pi)^3 e^2 n^{(2)}_{CI}}{\varepsilon_L^2 q_\perp^2} \quad ; \quad \left|\beta^{(a)}_{(CI)}\right| = \frac{2\pi^2 e^3 n^{(2)}_{CI}}{\varepsilon_L^2 \hbar v_F^2} \quad . \tag{3.7}$$

One can see from this formula that the values $\beta^{(a)}_{(CI)}$ are equal for electrons and holes and do not depend on temperature and gate voltage. If two-dimensional set of charged impurities is separated from graphene on a distance $l$, where $l \cdot \max\{k_B T, eV_a\} > \hbar v_F$, then

$$\left|\beta^{(a)}_{(CI)}\right| = \frac{\pi e^3 n^{(2)}_{CI}}{\varepsilon_L^2 k_B T v_F l} \quad . \tag{3.8}$$

### 3.3. Longitudinal acoustic phonons

Due to the strong inequality $v_F \gg s$ we can use the form (*A*5) in the limit $s/v_F \to 0$. Then at $\hbar\omega \ll k_B T$

$$\langle \varphi^2_{ph} \rangle_{\omega,\vec{q}\perp} = \Xi_A^2 k_B T / e^2 \rho s^3 \quad . \tag{3.9}$$

With the help of Eqs. (2.10) and (3.9) we obtain:

$$\beta^{(a)}_{(ph)} = \frac{\Xi_A^2}{\pi^2 e \rho s^3} \left(\frac{k_B T}{\hbar v_F}\right)^4 \Omega(\kappa_a) \quad ; \quad \Omega(\kappa_a) = \frac{\int_0^\infty \kappa^5 \exp(\kappa - \kappa_a)[1 + \exp(\kappa - \kappa_a)]^{-2} d\kappa}{\int_0^\infty \kappa [1 + \exp(\kappa - \kappa_a)]^{-1} d\kappa} \quad . \tag{3.10}$$

Here $\Omega(0) \approx 142$ ; $\Omega(\kappa_a \gg 1) \approx 2\kappa_a^4$. Therefore, for neutral graphene $\beta^{(a)}_{(ph)} \propto T^4$, for monopolar grapheme $\beta^{(a)}_{(ph)} \propto T^0 V_a^4$.

### 3.4. Drag-factor

For neutral graphene we use for calculations the Eq. (2.9) directly. For two-layer graphene structure we use Eq. (2.9), modified by the factor $\exp(-2q_\perp l)$ under the sign of integral over $\vec{q}_\perp$. Here $l$ is distance between the layers. Assume $l \cdot \max\{k_B T, eV_a\} > \hbar v_F$. In monopolar graphene the mutual drag of electrons and holes is negligible. For neutral graphene

$$\xi^{(h,e)} = -\xi^{(e,h)} = \frac{2\varpi e^3 (k_B T)^2}{\hbar^3 v_F^4 \varepsilon_L^2} \frac{\int_0^\infty t^2 \exp(t)[1+\exp(t)]^{-2} dt \int_0^\infty t \exp(t)[1+\exp(t)]^{-2} dt}{\int_0^\infty t[1+\exp(t)]^{-1} dt} \approx 1.386 \frac{e^3 (k_B T)^2}{\hbar^3 v_F^4 \varepsilon_L^2} \quad . \tag{3.11}$$

Here $\varpi = \int_0^\infty t^2 dt / \sinh^2(t) \approx 3.29$. For two-layer graphene

$$\xi^{(h,e)} = -\xi^{(e,h)} (V_e / V_h)^2 \approx 0.15 e^3 k_B T / \hbar^2 v_F^3 \varepsilon_L^2 l \quad . \tag{3.12}$$

## 4. Conductivity of neutral graphene

Here we assume $V_e = V_h = 0$. Then $\beta^{(h)} = -\beta^{(e)} = \beta > 0$, $\xi^{(h,e)} = -\xi^{(e,h)} = \xi > 0$, and one obtains from Eqs. (2.14) and (2.15) the system of equations:

$$\vec{E}_\perp + \beta \vec{u}^{(e)} + \xi(\vec{u}^{(e)} - \vec{u}^{(h)}) = 0 \ ; \tag{4.1}$$

$$\vec{E}_\perp - \beta \vec{u}^{(h)} - \xi(\vec{u}^{(h)} - \vec{u}^{(e)}) = 0 \ . \tag{4.2}$$

It follows from these expressions, that for negligible drag ($\xi \to 0$) the value $\beta$ is inverse mobility of carriers. For monopolar graphene one has good reason to neglect the drag, then the mobility $\mu^{(e,h)} = 1/\left|\beta^{(e,h)}\right|$.

Solving the system (4.1), (4.2), we find:

$$\vec{u}_\perp^{(h)} = -\vec{u}_\perp^{(e)} = \frac{1}{\beta + 2\xi} \vec{E}_\perp = \mu \vec{E}_\perp . \tag{4.3}$$

The density of the total current:

$$\vec{j}_\perp = \vec{j}_\perp^{(e)} + \vec{j}_\perp^{(h)} = \sigma_\perp \vec{E}_\perp = \frac{2en^{(e)}}{\beta + 2\xi} \vec{E}_\perp = 2en^{(e)} \mu \vec{E}_\perp \approx \frac{e}{\beta + 2\xi} \left(\frac{k_B T}{\hbar v_F}\right)^2 \vec{E}_\perp \ . \tag{4.4}$$

It is easy to see that for neutral graphene the contribution of electron-hole drag in the mobility $\mu$ is determined by the ratio $2\xi/\beta$.

If external scattering system is created by neutral impurities, then $2\xi/\beta \approx 0.6 q_B^2 / n_{(NI)}^{(2)}$. For charged impurities we have $2\xi/\beta \approx 0.9 (k_B T)^2 / \hbar^2 v_F^2 n_{(CI)}^{(2)}$. For LA-phonons $2\xi/\beta \approx 1.3 e^4 \rho s^3 / \Xi_A^2 \hbar v_F^2 \varepsilon_L^2$.

### 4. Conductivity of two-layer graphene

Two-layer graphene systems are especially interesting. That contain two parallel layers of graphene separated by ultrathin highly insulating dielectric layer (see, as possible example, two-gate composition on Fig. 1).

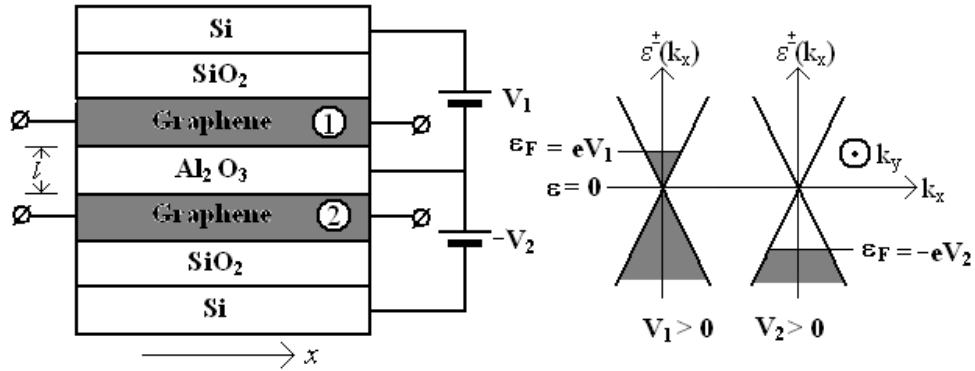

Fig. 1. Controlled two-layer graphene composition.

The gate voltages $V_{1,2}$ assume sufficiently high and temperature $T$ so low, that carriers in *1*- and *2*-graphene can be accepted as high-degenerate ($k_B T \ll eV_{1,2}$). Then in consequence with Eq. (*I*.5) the densities of carriers in separate layers are

$$n_{e,h} = \pi^{-1}(eV_{1,2}/v_F \hbar)^2 . \tag{5.1}$$

Represent for an example the charged impurities, disposed in the same planes as graphene, and consider here the most actual case $V_1 = V_e > 0$ and $V_2 = V_h > 0$. Then graphene-1 has electron conductivity and graphene-2 has hole conductivity.

At high gate voltages the system of equations (2.14) and (2.15) can be written in the following form (here $\beta^{(h)} > 0$; $\beta^{(e)} < 0$; $\xi^{(e,h)} < 0$):

$$E_{1x} - \beta^{(e)} u_x^{(e)} - \xi^{(e,h)} (u_x^{(e)} - u_x^{(h)}) = 0 \quad ; \tag{5.3}$$

$$= E_{2x} - \beta^{(h)} u_x^{(h)} - \xi^{(e,h)} (u_x^{(e)} - u_x^{(h)})(V_e / V_h)^2 = 0. \tag{5.4}$$

The electric field components $E_{1x}$ and $E_{2x}$ can have as the same signs as opposite that; both $V_{e,h} > 0$. Then solution of the system (5.3) and (5.4) has the form (here $-\xi^{(e,h)} = \xi > 0$)

$$u_x^{(e)} = \frac{E_{1x}\beta^{(h)} - \xi[E_{2x} - (V_e/V_h)^2 E_{1x}]}{\beta^{(e)}\beta^{(h)} + \xi[(V_e/V_h)^2 \beta^{(e)} - \beta^{(h)}]} \quad ; \quad u_x^{(h)} = \frac{-E_{2x}\beta^{(e)} + \xi[E_{2x} - (V_e/V_h)^2 E_{1x}]}{\beta^{(e)}\beta^{(h)} + \xi[\beta^{(h)} + (V_e/V_h)^2 \beta^{(e)} - \beta^{(h)}]} . \tag{5.5}$$

It follows from Eqs. (5.1) and (5.5):

$$j_x^{(n)} = -e n_e u_x^{(e)} = \frac{e}{2\pi} \left(\frac{eV_e}{v_F \hbar}\right)^2 \frac{E_{1x}\beta^{(h)} - \xi[E_{2x} - (V_e/V_h)^2 E_{1x}]}{-\beta^{(e)}\beta^{(h)} + \xi[\beta^{(h)} - (V_e/V_h)^2 \beta^{(e)}]} \quad ; \tag{5.6}$$

$$j_x^{(p)} = e n_h u_x^{(h)} = \frac{e}{2\pi} \left(\frac{eV_h}{v_F \hbar}\right)^2 \frac{-E_{2x}\beta^{(e)} + \xi[E_{2x} - (V_e/V_h)^2 E_{1x}]}{-\beta^{(e)}\beta^{(h)} + \xi[\beta^{(h)} - (V_e/V_h)^2 \beta^{(e)}]} . \tag{5.7}$$

One can see from these expressions that both currents $j_x^{(n)}$ and $j_x^{(p)}$ are controlled by two forced fields, $E_{1x}$ and $E_{2x}$, and two gate voltages, $V_e$ and $V_h$.

If the scattering by charged impurities excels that, related to neutral centers and *LA*-phonons, we can suppose (see Eq. (5.2))

$$-\beta^{(e)} = \beta^{(h)} = \beta . \tag{5.8}$$

Then the current densities can be represented in the forms

$$j_x^{(n)} = -e n_e u_x^{(e)} = \sigma^{(n)} E_{1x} = (e/2\pi\beta)(eV_1/v_F\hbar)^2 D^{(n)} E_{1x}; \tag{5.9}$$

$$j_x^{(p)} = e n_h u_x^{(h)} = \sigma^{(p)} E_{1x} = (e/2\pi\beta)(eV_2/v_F\hbar)^2 D^{(p)} E_{2x}. \tag{5.10}$$

Here $D^{(n)}$ and $D^{(p)}$ are drag-factors, responsible for the electron-hole drag in considered two-layer graphene structure. They are

$$D^{(n)} = \frac{(\beta/\xi) - [(E_{2x}/E_{1x}) - (V_1/V_2)^2]}{(\beta/\xi) + [1 + (V_1/V_2)^2]} ; \quad D^{(p)} = \frac{(\beta/\xi) - [1 - (V_1/V_2)^2 (E_{1x}/E_{2x})]}{(\beta/\xi) + [1 + (V_1/V_2)^2]} \tag{5.11}$$

In the absence of a drag ($\xi \to 0$) we have: $D^{(n)} = D^{(p)} = 1$. The symmetry of conductivities $\sigma^{(n)}$ and $\sigma^{(p)}$ relatively to the permutation $1 \leftrightarrow 2$ is evident. So it is enough to investigate the structure of one drag-factor.

Below Figs. 2 – 4 illustrate dependence of the drag-factor $D^{(n)}$ on the value of three controlling combinations. If the factor $D^{(n)}$ changes the sign, it means that the drag of electrons by holes reverses the direction of electron current.

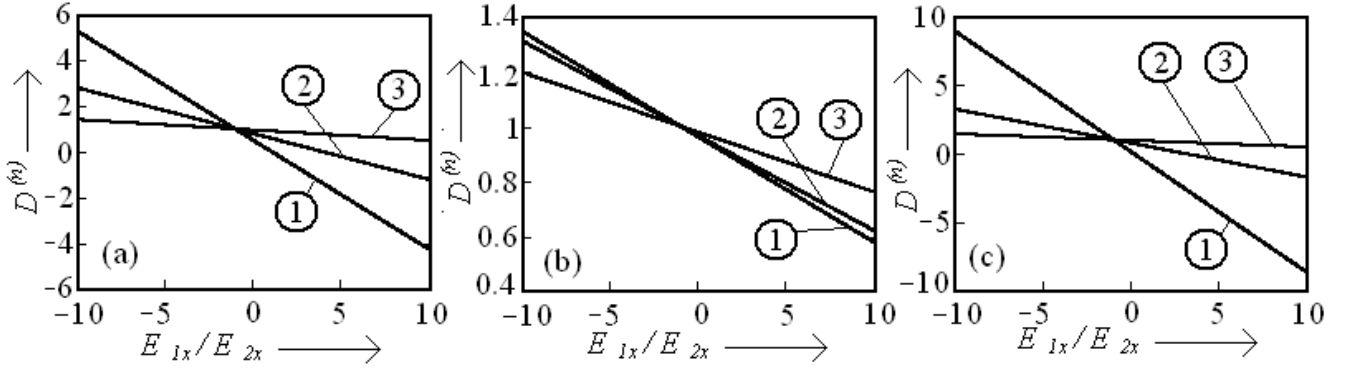

Fig. 2. Dependence of the drag-factor $D^{(n)}$ on the ratio $E_{1x}/E_{2x}$. a) – $V_1/V_2 = 1$ ; b) – $V_1/V_2 = 5$ ; c) – $V_1/V_2 = 0.2$ ;
1) – $\beta/\xi = 0.1$ ; 2) – $\beta/\xi = 3$ ; 3) – $\beta/\xi = 20$ .

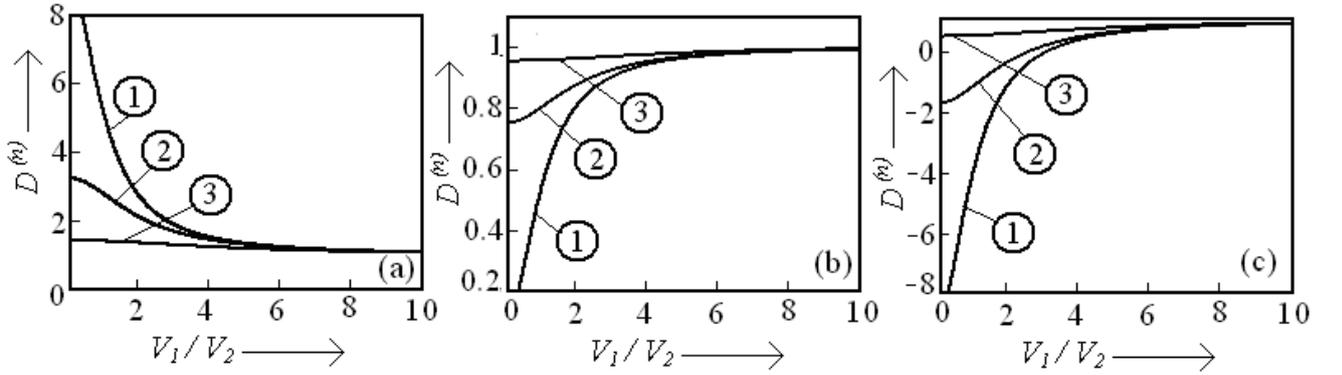

Fig. 3. Dependence of the drag-factor $D^{(n)}$ on the ratio $V_1/V_2$ .a) – $E_{1x}/E_{2x} = -10$; b) – $E_{1x}/E_{2x} = 0$; c) – $E_{1x}/E_{2x} = 10$;
1) – $\beta/\xi = 0.1$ ; 2) – $\beta/\xi = 3$ ; 3) – $\beta/\xi = 20$ .

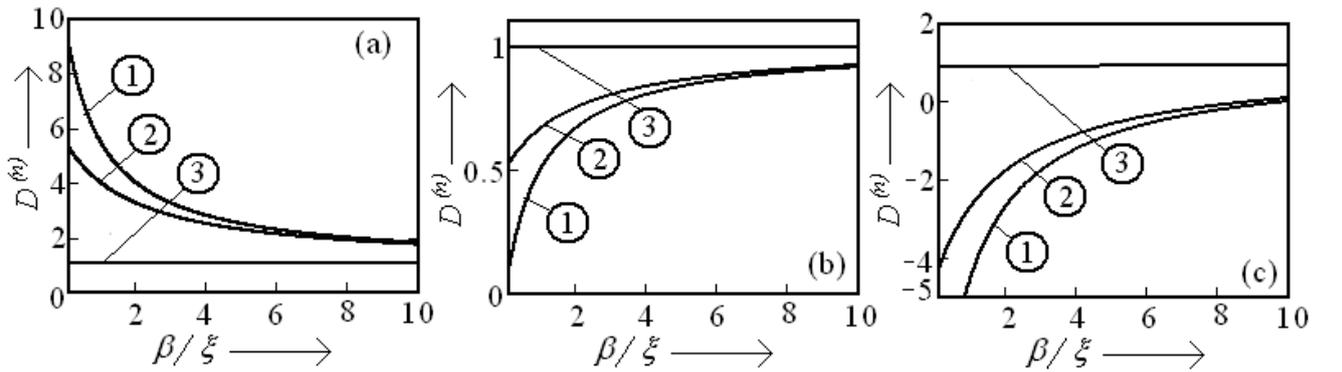

Fig. 4. Dependence of the drag-factor $D^{(n)}$ on the ratio $\beta/\xi$ .a) – $E_{1x}/E_{2x} = -10$; b) – $E_{1x}/E_{2x} = 0$; c) – $E_{1x}/E_{2x} = 10$;
1) – $V_1/V_2 = 0.1$ ; 2) – $V_1/V_2 \to 0$ ; 3) – $V_1/V_2 = 10$ .

**Appendix**

As the external scattering system we consider here impurities, distributed uniformly in plane $z = 0$ with the density $n_I^{(2)}$, and *LA*-phonons. To calculate the two-dimensional correlator related to the plane $z = 0$, we use the formula

$$\langle \varphi^2 \rangle_{\omega, \vec{q}\perp} = \frac{1}{2\pi} \int_{-\infty}^{\infty} \langle \varphi^2 \rangle_{\omega, \vec{q}} dq_z \qquad (A1)$$

and different necessary formulae from Refs. [5] and [6].

The three-dimensional potential of one "neutral" hydrogen-like impurity is (see Ref. [7])

$$\varphi_{NI}(\vec{r}) = (e_I / \varepsilon_L r)(1 + rq_B / 2) \exp(-rq_B). \qquad (A2)$$

Here $q_B = 2/r_B = 2m_0 e^2 / \varepsilon_L \hbar^2$.

After Fourier-transformation one obtains:

$$\varphi_{NI}(\vec{q}, \omega) = (4\pi e_I / \varepsilon_L)(q^2 + q_B^2)^{-1}[1 + q_B^2 /(q^2 + q_B^2)]\delta(\omega);$$

$$\varphi_{NI}(\vec{q}_\perp, \omega) = (2\pi e_I / \varepsilon_L)(q^2 + q_B^2)^{-1/2}[1 + q_B^2 / 2(q^2 + q_B^2)]\delta(\omega).$$

At $q_B \gg q_\perp$

$$\langle \varphi_{NI}^2 \rangle_{q_\perp} = 9\pi e^2 n_{NI}^{(2)} / 2\varepsilon_L^2 q_B^2. \qquad (A3)$$

For charged impurities

$$\varphi_{CI}(\vec{r}) = \frac{e_I}{\varepsilon_L r} \; ; \quad \varphi_{CI}(\omega, \vec{q}) = \frac{4\pi e_I}{\varepsilon_L q^2} \delta(\omega) \; ; \quad \langle \varphi_{CI}^2 \rangle_{\vec{q}\perp} = \frac{(2\pi)^3 e^2 n_{CI}^{(2)}}{\varepsilon_L^2 q_\perp^2} . \qquad (A4)$$

For *LA*-phonons with dispersion law $\omega = sq$ and for the deformation potential $\Xi_A$

$$\langle \varphi_{ph}^2 \rangle_{\omega, \vec{q}} = \left(\frac{\Xi_A}{e}\right)^2 \frac{\pi \hbar \omega}{2\rho s^2} \coth\left(\frac{\hbar \omega}{2k_B T}\right) [\delta(\omega - sq) + \delta(\omega + sq)] \; ;$$

$$\langle \varphi_{ph}^2 \rangle_{\omega, \vec{q}\perp} = \left(\frac{\Xi_A}{e}\right)^2 \frac{\hbar \omega^2}{2\rho s^3} \coth\left(\frac{\hbar |\omega|}{2k_B T}\right) \frac{\vartheta(\omega^2 - s^2 q_\perp^2)}{\sqrt{\omega^2 - s^2 q_\perp^2}} . \qquad (A5)$$

Here $s$ is the sound velocity and $\vartheta(\omega^2 - s^2 q_\perp^2)$ is the stepped function.

[1]. Castro Nero *et al*. The electronic properties of graphene. // *Rev. Mod. Phys.* **81**, p.109 (2009).

[2]. P. R. Wallace.// Phys. Rev. **71**, p.622 (1947).

[3]. K. S. Novoselov *et al*. // Two-dimensional gas of massless Dirac fermions in graphene. // *Nature*, **438**, p.197 (2005).

[4]. S.Kim *et al.* Coulomb drag of massless fermions in graphene. // arXiv. 1010.2113v1 [cond.-mat. mes-hall] 11 Oct 2010.

[5]. I. I. Boiko: *Transport of Carriers in Semiconductors*, Ed. V. Lashkaryov Inst. of Semiconductor Physics, NAS of Ukraine, Kyiv, 2009 (in Russian).

[6]. I. I. Boiko: *Kinetics of electron gas interacting with fluctuating potential*, Naukova dumka, Kiev (1993) (in Russian).

[7]. I. I. Boiko: Impurity scattering of band carriers. // *Semiconductor Physics, Quantum Electronics & Optoelectronics* (to be published).